\documentstyle[12pt,epsfig]{article}

\def\bc{\begin{center}}
\def\ec{\end{center}}
\def\beq{\begin{equation}}
\def\eeq{\end{equation}}

\def\at#1{\left. \right|^{}_{#1}}
\def\hs#1{\hspace*{#1cm}}

\def\av#1{\langle {#1} \rangle}
\def\avr#1#2{\langle {#1} \rangle^{}_{#2}}
\def\etai{\sqrt{\eta_i}}
\def\F{{n^{}_F}}
\def\B{{n^{}_B}}

\def\ppB{{p^{2}_{{\rm t}B}}}
\def\Ceta{{\{\eta^{}_1,...,\eta^{}_M\}}}
\def\Aeta{{\eta^{}_1,...,\eta^{}_M}}
\def\Cn{{\{n^{}_1,...,n^{}_M\}}}

\def\ol#1{\overline{#1}}

\def\reta{\sum_{i}^{} \sqrt{\eta_i}}
\def\seta{\sum_{i}^{} \eta_i}
\def\ski{\sum_{i}^{} k_i}
\def\sni{\sum_{i}^{} n_i}

\def\bfs{{\bf s}}
\def\olb{\overline{b}}

\def\olx{\overline{x}}

\def\bN{\overline{N}}

\def\bn{\overline{n}}

\def\beti{\ol{\eta}^{}_{i}}

\def\j#1#2#3#4{{#1} {\bf #2} (#3) #4}

\def\NP{Nucl. Phys.}
\def\PL{Phys. Lett.}
\def\PRL{Phys. Rev. Lett.}

\def\ZP{Z. Phys.}
\def\PRep{Phys. Rep.}

\def\EPJ{Eur. Phys. J.}
\def\IJMP{Int. J. of Mod. Phys.}

\title{
Simple Cellular Model of\\ 
Long-Range Multiplicity and $p_t$ Correlations\\
in High-Energy Nuclear Collisions
}
\author{
{\bf V.V. Vechernin and R.S. Kolevatov}\\
{\it High Energy Physics Dep., St. Petersburg State University,}\\
{\it 1 Ulianovskaya str., 198504 St.Petersburg, Russia}
}
\date{}

\begin{document}

\begin{titlepage}
\maketitle
\medskip

%  Abstract  XXXXXXXXXXXXXXXXXXXXXXXXXXXXXXXXXXXXXXXXXXXXXXXXXXXXXX

\begin{abstract}
A simple cellular model for the description of
the long-range multiplicity and $p_t$ correlations
in high-energy nuclear collisions
originating from the string fusion model is proposed.
Three versions of the model: without fusion,
with local and with global string
fusion are formulated.

A Gauss approximation which
enables explicit analytical calculations
of the correlation functions in some asymptotic cases
in the framework of the model is developed.
The assumptions of the model
and the validity of a Gauss approximation
are checked up in the simplest (no fusion)
case when the explicit solution of the model can be found.

The role of the size of cells is anylised.
The modification of the results in the case of non-Poissonian
distributions is also discussed.
\end{abstract}
\vspace*{1cm}
\end{titlepage}

%  1  XXXXXXXXXXXXXXXXXXXXXXXXXXXXXXXXXXXXXXXXXXXXXXXXXXXXXX

\section{Introduction.}

The colour strings approach \cite{Capella1,Kaidalov}
is widely applied for the description of the soft part of
the hadronic and nuclear interactions at high energies.

In the frame work of this approach
the string fusion model was suggested \cite{BP1}.
Later it was developed \cite{ABP1}-\cite{BPR} and applied for the
description of the long-range multiplicity and $p_t$ correlations
in relativistic nuclear collisions \cite{PRL94}-\cite{BP00}.

The aim of the present paper is to formulate some simple cellular
analog of the model, which enables explicit analytical calculations
of the correlation functions in some asymptotic cases and
drastically simplifies calculations in the case of real nucleus
collisions.

We check up the assumptions of the cellular model
and the validity of a suggested Gauss approximation
in the simplest (no fusion)
case when the explicit solution of the model can be found.

The paper organized as follows.
Next section is devoted to the formulation
of the a cellular analog of the string fusion model.
The version of the model with a local string
fusion is considered.

The section 3 deals with the no string fusion limit of the model.
The correspondence with the previous results \cite{BPV00} is demonstrated.

In the section 4 a Gauss approximation for the
correlation function calculations is formulated.
The results of the calculations in this approximation
is compared with exact solution,
which can be found in the no fusion case.

In the section 5 the role of the size of cells is anylised.
The cluster size dependence is considered
and the version of the model with global string
fusion is formulated.

The modification of the results in the case of non-Poissonian
distributions is discussed in the section 6.

%  2  XXXXXXXXXXXXXXXXXXXXXXXXXXXXXXXXXXXXXXXXXXXXXXXXXXXXXX

\section{Cellular approach to the string fusion phe\-no\-me\-non.}

Let us consider the collision of nuclei in two stage scenario
when at first stage the colour strings are formed,
and at the second stage these strings 
(or some other (higher colour) strings
formed due to fusion of primary strings) are decaying, emitting
observed particles.

We'll consider three possibilities:
without string fusion,
with local and with global string fusion.
The case with a local fusion corresponds to the model,
where colour fields
are summing up only locally and the global fusion case corresponds
to the model, where colour fields are summing up globally -
all over the cluster area - into one average colour field,
the last case corresponds to the summing of the sources colour charges.
(In section 5 we are refering to these cases as A) and B) correspondingly.)

In the transverse plane depending on the impact
parameter $b$ we have some interaction area $S(b)$.
Let us split this area on the cells
of order of the transverse string size.
Then we'll have $M=S(b)/\sigma_0$ cells,
where $\sigma_0=\pi r_0^2$ is the transverse square of the string
and $r_0\approx 0.2 fm$ is the string radius.

{\bf Local string fusion.}
At first let us consider the case with a local fusion.
In this case the assumption of the model is 
that if the number of strings
belonging to the $i$-th cell is $\eta_i$, then they form
higher colour string, which emits in average $\mu _0\sqrt{\eta _i}$ particles
with mean $p_t^2$ equal to $p^2\sqrt{\eta _i}$,
compared with $\mu _0$ particles with $\langle p_t^2 \rangle =p^2$
emitting by a single string.
(Note that situation when some $\eta_i=0$ is also admitted.)

Let us denote by $n_i$ and $\bn_i$ - the number and the average
number of particles emitted by the higher string from $i$-th cell
in a given rapidity interval, then
\beq
\bn_i=\mu _0\sqrt{\eta _i}
\label{fus1}
\eeq

From event to event the number of strings $\eta_i$ in $i$-th cell
will fluctuate around some
average value - $\overline{\eta }_i^{}$.
Clear that in the case of real nuclear collisions
these average values $\overline{\eta }_i^{}$ will be different
for different cells. They will depend on the position (${\bf s}$)
of the $i$-th cell in the interaction area
(${\bf s}$ is two dimensional vector in transverse plane).
To get the physical answer we have to sum the contributions
from different cells, which corresponds to integration over
${\bf s}$ in transverse plane.

The average local density of primary strings $\overline{\eta }_i^{}$
in the point ${\bf s}$ of transverse plane is uniquely
determined by the distributions of nuclear densities and
the value of the impact parameter - $b$.
They can be calculated, for example, in Glauber approximation.
We'll do this later in a separate paper. In present paper
we consider that all $\overline{\eta }_i^{}$ are already
fixed from these considerations at given value 
of the impact parameter - $b$.

If we introduce:
\beq
N=\sum_{i=1}^M \eta_i , \hs{1} \bN=\sum_{i=1}^M \beti
\label{not}
\eeq
then clear that $N$ is the number of strings in the given event
and $\bN$ is the mean number of strings for this type of events
(at the fixed impact parameter $b$.

To go to long-range rapidity correlations let us consider
two rapidity windows $F$ (forward) and $B$ (backward).
Each event corresponds to a certain configuration $\Ceta$ of
strings and certain numbers of charged particles $\Cn$ emitted by
these strings in the forward rapidity window. Then the total
number of particles produced in the forward rapidity window
will be equal to $n_F$:
\beq
n_F=\sum_{i=1}^{M} n_i
\label{nF}
\eeq
The probability to detect
$n_F$ particles in the forward rapidity window
for a given configuration $\Ceta$ of strings
is equal to
\beq
P^{}_{\{\eta^{}_1,...,\eta^{}_M\}} (\F)=
\sum_{\{n^{}_1,...,n^{}_M\}} \delta_{\F,\sum_i n_i} \prod_{i=1}^M
p^{}_{\eta_i} (n_i)
\label{PnF}
\eeq
where $p^{}_{\eta_i} (n_i)$ is the probability of the emission
of $n_i$ particles by the string $\eta_i$ in the forward
rapidity window. By our assumption (\ref{fus1})
\beq
\bn_i\equiv\sum_{n_i=0}^{\infty} n_i p^{}_{\eta_i} (n_i) =
\mu_0 \sqrt{\eta_i}
\label{fusion}
\eeq

If we denote else by $W(\Aeta)$ the probability of realization of the
string configuration $\Ceta$ in the given event,
then the average value of some quantity $O$ under condition of
the production of $n_F$ particles in the forward window
will be equal to
\beq
\avr{O}{\F}
=\frac{\sum_{\{\eta^{}_1,...,\eta^{}_M\}} \avr{O} {{\{\eta^{}_1,...,\eta^{}_M\}},\F} W(\Aeta) P^{}_{\{\eta^{}_1,...,\eta^{}_M\}} (\F)}
{\sum_{\{\eta^{}_1,...,\eta^{}_M\}} W(\Aeta) P^{}_{\{\eta^{}_1,...,\eta^{}_M\}} (\F)}
\label{avO}
\eeq
One has to omit in this $M$-fold sums one term, when all $\eta_i=0$,
which corresponds to the absence of inelastic interaction between
the nucleons of the colliding nuclei.

If the $O$ in the number of particles produced in the backward
rapidity window $n_B$ in the given event, then
(for $\avr{\B}{\F}$ correlations) we have to use
\beq
\avr{\B} {{\{\eta^{}_1,...,\eta^{}_M\}},\F} =
\mu_0 \sum_{i=1}^{M} \sqrt{\eta_i}
\label{Onn}
\eeq
If the $O$ in the mean squared transverse momentum
of particles produced in the backward
rapidity window $\ppB$ in the given event, then
(for $\avr{\ppB}{\F}$ correlations) we have to use:
\beq
\avr{\ppB} {{\{\eta^{}_1,...,\eta^{}_M\}},\F}
= \sum_{i=1}^{M} \frac{\sqrt{\eta_i}}
{\sum_{i=1}^{M} \sqrt{\eta_i}} p^2 \sqrt{\eta_i}
= p^2 \frac{\sum_{i=1}^{M} \eta_i}{\sum_{i=1}^{M} \sqrt{\eta_i}}
\label{Optn}
\eeq

Later we'll assume that numbers of primary strings in each cell $\eta_i$
fluctuate independently around some average quantities
$\overline{\eta }_i^{}$ uniquely
determined by the distributions of nuclear densities and
the value of the impact parameter - $b$ (see above). then
\beq
W(\Aeta) = \prod_{i=1}^M w (\eta_i), \hs{1}
\sum_{i=1}^M \eta_i w (\eta_i) = \beti
\label{WCeta}
\eeq

For clearness we'll sometimes address to a simple
"homogeneous" case, when all $\beti$
(but not the $\eta_i$, which fluctuate!)
is equal each other in the interaction area $\beti=\eta$.
The parameter $\eta$ coincides in this case with
the parameter $\eta$ used in the papers \cite{BPR,BPep00,BP00}
and has the meaning of the mean number of strings
per area of one string ($\eta=({\rm mean\ string\ density})\times\sigma_0$).
In general case the parameters $\beti$ have the same meaning,
but with mean string density depending on the point $\bfs$
in the transverse interaction plane
($\beti=({\rm mean\ string\ density\ in\ the\ point}\ \bfs)\times\sigma_0$).

If we assume else the Poissonian form of $p^{}_{\eta_i} (n_i)$
($\rho^{}_{a}(x)$ is the Poisson distribution with $\olx=a$):
\beq
p^{}_{\eta_i} (n_i) =  \rho^{}_{\mu_0\etai}(n_i)
 \equiv e^{-\mu_0\etai}\frac{(\mu_0\etai)^{n_i}}{n_i!}
\label{Poisson}
\eeq
then we find for $P^{}_{\{\eta^{}_1,...,\eta^{}_M\}} (\F)$:
\beq
P^{}_{\{\eta^{}_1,...,\eta^{}_M\}} (\F) = \rho^{}_{\mu_0\reta}(\F)
\label{Poisso}
\eeq

%  3  XXXXXXXXXXXXXXXXXXXXXXXXXXXXXXXXXXXXXXXXXXXXXXXXXXXXXX

\section{No string fusion. Correspondence with the previous results.}

{\bf No string fusion.}
In the no fusion case we have the same formulae, but instead of
(\ref{fusion}), (\ref{Onn}) and (\ref{Optn}), we have to use
\beq
\bn_i=\sum_{n_i=0}^{\infty} n_i p^{}_{\eta_i} (n_i) = \mu_0 \eta_i
\label{nf}
\eeq
\beq
\avr{\B} {{\{\eta^{}_1,...,\eta^{}_M\}},\F} = \mu_0 \sum_{i=1}^{M} \eta_i
\label{nf1}
\eeq
\beq
\avr{\ppB} {{\{\eta^{}_1,...,\eta^{}_M\}},\F}
= \sum_{i=1}^{M} \frac{\eta_i}{\sum_{i=1}^{M} \eta_i} p^2
= p^2
\label{nf2}
\eeq
which immediately leads to the absence of $\avr{\ppB}{\F}$ correlations.

In this case we have to use also
\beq
p^{}_{\eta_i} (n_i) =  \rho^{}_{\mu_0\eta_i}(n_i)
 \equiv e^{-\mu_0\eta_i}\frac{(\mu_0\eta_i)^{n_i}}{n_i!}
\label{nfpois}
\eeq
and
\beq
P^{}_{\{\eta^{}_1,...,\eta^{}_M\}} (\F) = \rho^{}_{\mu_0\seta}(\F)
\label{nfPois}
\eeq
instead of (\ref{Poisson}) and (\ref{Poisso}).
Then for $\avr{\B}{\F}$ correlations we find
\beq
\avr{\B}{\F}
=\frac{\mu_0\sum_{\{\eta^{}_1,...,\eta^{}_M\}} \left(\seta\right) \left(\prod_i w (\eta_i)\right)
\rho^{}_{\mu_0\seta}(\F)}
{\sum_{\{\eta^{}_1,...,\eta^{}_M\}} \left(\prod_i w (\eta_i)\right) \rho^{}_{\mu_0\seta}(\F)}
\label{BF1}
\eeq
Introducing under the sums
\beq
1=\sum_N \delta_{N,\seta}
\label{1}
\eeq
and putting everywhere $\seta=N$ we find
\beq
\avr{\B}{\F}
=\frac{\mu_0\sum_N N W(N) \rho^{}_{\mu_0 N}(\F)}
        {\sum_N W(N) \rho^{}_{\mu_0 N}(\F)}
\label{BF2}
\eeq
where
\beq
W(N) = \sum_{{\{\eta^{}_1,...,\eta^{}_M\}}} \delta_{N,\seta} \prod_i w (\eta_i)
\label{WN}
\eeq
If we also admit the Poissonian form for $w (\eta_i)$
and that all $\beti=\eta$ (the homogeneous case):
\beq
w (\eta_i) =  \rho^{}_{\eta}(\eta_i)
\equiv e^{-\eta}\frac{(\eta)^{\eta_i}}{\eta_i!}
\label{wpois}
\eeq
then we find
\beq
W (N) =  \rho^{}_{M\eta}(N)
\label{WNpois}
\eeq
and (\ref{BF2}) has the form
\beq
\avr{\B}{\F}
=\frac{\mu_0\sum_N N \rho^{}_{M\eta}(N) \rho^{}_{\mu_0 N}(\F)}
        {\sum_N \rho^{}_{M\eta}(N) \rho^{}_{\mu_0 N}(\F)}
\label{BF3}
\eeq
As $\bN \equiv \sum_N N W(N)= M\eta$ we see, that (\ref{BF3})
coincides with the formula from the paper \cite{BPV00},
where the notation $\bn=\mu_0$ for the mean multiplicity
from one string (emitter). The $N$ and $\bN$ is the number and
the mean number of strings (emitters) was used.
\beq
\avr{\B}{\F}
=\frac{\bn\sum_N N \rho^{}_{\bN}(N) \rho^{}_{\bn N}(\F)}
        {\sum_N \rho^{}_{\bN}(N) \rho^{}_{\bn N}(\F)}
\label{BF4}
\eeq

%  4  XXXXXXXXXXXXXXXXXXXXXXXXXXXXXXXXXXXXXXXXXXXXXXXXXXXXXX

\section{Gauss approximation.}

Let us now evaluate (\ref{BF4}) at $\bN\gg 1$ and $\mu_0\bN\gg 1$.
At these assumptions we can replace $\sum_N$ by $\int dN$ and
the Poissonian distributions by Gaussian distributions
($g^{}_{a, \sigma}(x)$ is the Gauss distribution with $\olx=a$
and $\overline{x^2}-\olx^2=\sigma^2$).
\beq
\rho^{}_{\mu_0 N}(\F)
\rightarrow g^{}_{\avr{\F}{N}, \sigma_F}(\F)
\equiv \frac{1}{\sqrt{2\pi}\sigma_F}e^{-\frac{(\F-\avr{\F}{N})^2}{2\sigma_F^2}}
\label{pgaus}
\eeq
with  $\sigma_F^2=\avr{\F}{N}=\mu_0 N$.
Similarly
\beq
\rho^{}_{\bN}(N)
\rightarrow g^{}_{\bN, \sigma_N}(N)
\equiv \frac{1}{\sqrt{2\pi}\sigma_N}e^{-\frac{(N-\bN)^2}{2\sigma_N^2}}
\label{wgaus}
\eeq
with  $\sigma_N^2=\bN$.

So we find
\beq
\avr{\B}{\F}
=\frac{\mu_0\int dN N \frac{1}{\sqrt{N}}e^{-\varphi(N,\F)}}
{\int dN \frac{1}{\sqrt{N}}e^{-\varphi(N,\F)}}
\label{BF5}
\eeq
with
\beq
\varphi(N,\F)=\frac{(N-\bN)^2}{2\bN}+\frac{(\F-\mu_0 N)^2}{2\mu_0 N}
\label{phi}
\eeq
and
\beq
\frac{d\varphi}{dN}=\frac{N}{\bN}-1+\frac{\mu_0}{2}-\frac{\F^2}{2\mu_0 N^2}
\label{dphi}
\eeq
Let us denote by $N_*$ the point where  $\frac{d\varphi}{dN}=0$.
Then we can evaluate (\ref{BF5}) as follows
\beq
\avr{\B}{\F} =\mu_0 N_* (\F)
\label{BF6}
\eeq
In relative variables we can rewrite condition $\frac{d\varphi}{dN}=0$
as follows
\beq
z^3-z^2=\frac{\mu_0}{2}(f^2-z^2)
\label{z}
\eeq
which defines $z$ as function of $f$,
where $z=N_*/\bN$ and $f=\F/\av{\F}=\F/(\mu_0 \bN)$ and then
\beq
\avr{\B}{\F} =\mu_0 N_* (\F)=\mu_0 \bN z(f)=\av{\F} z(f)
\label{BF7}
\eeq
Let us define correlation coefficient as
\beq
b\equiv\frac{d\avr{\B}{\F}}{d\F} \at {\F=\av{\F}}
=\frac{dz}{df} \at {f=1}
\label{b1}
\eeq
From (\ref{z}) we have
\beq
\frac{dz}{df}=\frac{\mu_0 f}{3z^2+z(\mu_0-2)}
\label{dz}
\eeq
and then
\beq
b=\frac{\mu_0}{\mu_0+1}
\label{b2}
\eeq
Because as clear from (\ref{z}) at $f=1$ one has $z=1$.

In reality for one string $\mu_0=\frac{d\mu}{dy}\Delta y$,
where $\frac{d\mu}{dy}\simeq1.0\div1.2$ and $\Delta y$ is
the width of the rapidity window.
If one chooses backward and forward windows of a different width
$\Delta y_B \neq \Delta y_F$,
then $\mu_{0B} \neq \mu_{0F}$, and instead of (\ref{b2}) we have
\beq
\frac{d\avr{\B}{\F}}{d\F} \at {\F=\av{\F}} \equiv b=\frac{\mu_{0B}}{\mu_{0F}+1}
\label{b3}
\eeq
or for "relative" quantities
\beq
\frac{d\avr{\B}{\F}/\av{\B}}{d\F/\av{\F}} \at {\F=\av{\F}}
\equiv \ol{b}=\frac{\mu_{0F}}{\mu_{0F}+1}
\label{b4}
\eeq
We see that in the last case correlation coefficient depends only
on value of $\mu_{0F}$ in the {\it forward} window 
(see physical explanation of this fact in the end of the section).

In Figs.1-2 we present the results of the exact (dashed lines)
and approximate (in the Gauss approximation)(solid lines) calculations
of the function $\avr{\B}{\F}$
using formulas (\ref{BF4}) and (\ref{BF7}) correspondingly
at the different values of the mean number of strings $\bN=M\eta$ and $\mu_0$
in the no fusion case. We see that the Gauss approximation works
very well starting from $\bN=4$ especially in the region $\F=\av{\F}$,
where the most of experimental points lay and where correlation
coefficient is defined.

One has also to pay attention on the approximate linearity
of the correlation functions obtained here 
in the case of two Poisson distributions
by use of the formula (\ref{BF4}), 
which coincides with the corresponding formula from \cite{BPV00},
the problem dealt with in that paper.

In Figs.3
we present the results of the exact (dashed lines)
and approximate (in the Gauss approximation)(solid lines) calculations
of the correlation coefficient
$b \equiv \frac{d\avr{\B}{\F}}{d\F} \at {\F=\av{\F}}$ as function of $\mu_0$
using formulas (\ref{BF4}) and (\ref{b2}) correspondingly
at the different values of the mean number of strings $\bN=M\eta$
in the no fusion case. We see again that Gauss approximation works
very well starting from $\bN=4$ for any $\mu_0$.

As one can see from Figs.1-3
we have $\bN$-independence for correlation functions and correlation
coefficient $\ol{b}$ starting very early (from $\bN=4$).
It's interesting to mention that as one can see from Figs.4,5
at that the resulting distributions $P(\F)$ are changing drastically
with $\bN$ from $\bN=4$ to $\bN=128$.

We see also that we have practically ideal Gauss distribution for $P(\F)$
at $\bN=128$. This is in agreement with the experimental data.
Unlike the case of pp-interactions (where NBD for $P(\F)$ takes place)
in the case of nuclear collisions with a large number of emitting centers
the ideal Gauss distribution for $P(\F)$ has been observed experimentally
for central PbPb collisions
(i.e. at {\it fixed} value of impact parameter $b$ and hence fixed $S(b)$)
(see, for example, Fig.6 in \cite{Heiselberg}).
Note that for nuclear collisions in Glauber approximation
we have also Gauss distribution for $P(\F)$ at {\it fixed} value of impact parameter $b$
(see, for example, \cite{NA57}).

{\bf Physical interpretation.}
Let us to discuss in the conclusion of the section why
the correlation coefficient for "relative" quantities
$\olb$ (\ref{b4}) depends only on the multiplicity in the
{\it forward} rapidity window.

The correlations between $\B$ and $\F$ in the model
under consideration 
arise only through fluctuations
in the number of strings $N$.
At large values of $\mu_{0F}\gg 1$ it's more probable
that fluctuation in the number of forward produced particles $\F$
was caused by the fluctuation in the number of strings $N$ than by
the fluctuations in $n_i$ and vice versa
at small values of $\mu_{0F}\ll 1$ it's more probable
that fluctuation in the number of forward produced particles $\F$
was caused not by the fluctuation in the number of strings $N$ but by
the fluctuations in $n_i$.

Formally we can see it from analysis of the maximum $N_*$ of the function
$\varphi(N,\F)$ (\ref{phi}), in which the first term originates from
the fluctuations of $N$ and the second term originates from
the fluctuations of $n_i$. If we have some fluctuation
(i.e. $\F \neq \mu_0 \bN$), then at $\mu_{0F}\gg 1$ \  $N_*\to \F/\mu_0$
and at $\mu_{0F}\ll 1$ \  $N_*\to \bN$.

So if every string emits in average
large number of particles ($\mu_{0F}\gg 1$)
in the given forward interval $\Delta y_F$
then based on information of $\F$
we can do the justified conclusion on the number of strings $N$
in the given event
and hence expect corresponding change in the $\B$.
And if every string emits in average
small number of particles ($\mu_{0F}\ll 1$)
in the given interval $\Delta y_F$
then based on information of $\F$
we can't do any conclusions on the number of string $N$
in the given event.

Note that more detail analysis shows that this conclusion is not
based on the specific (Poissonian) form of the distributions.
At small $\bn_i\ll 1$ one always will have $\sigma_i\gg \bn_i$
due to discrete nature of $n_i$.

%  5  XXXXXXXXXXXXXXXXXXXXXXXXXXXXXXXXXXXXXXXXXXXXXXXXXXXXXX

\section{Cell size, cluster size, global string fusion.}

Let us go back to the string fusion and
consider the case, when $r_c$ - correlation radius
(cluster size) is not equal to the string radius - $r_0$
($\gamma\equiv r_c/r_0>1$), then the square of the cluster
in a transverse plane will be equal to $\Delta S=\gamma^2\sigma_0$
and $\gamma^2$ is the square of the cluster in string square units.
If j-th cluster was formed in the given event by $m_j$ strings,
then it will have $\eta_{cj}=m_j \sigma_0/\Delta S=m_j/\gamma^2 $,
where $j=1,...,M_c$ and $M_c$ is the number of the clusters:
$M_c=S(b)/\Delta S= \gamma^2)=M/\gamma^2$.
(We keep notation $M$ for the quantity $S(b)/\sigma_0=M$.)
The $S(b)$ is the transverse square of interaction area of nuclei at given
impact parameter $b$.
Note that in given approach we consider the clusters of the fixed (by hands)
area ($\Delta S$) with fluctuating from event to event number
of strings $m_j$ forming it
(i.e. the $\eta_{cj}$ fluctuates around $\eta$ in the homogeneous
case).

Now the elementary emitters will be not strings, but clusters.
The mean number of particles emitted by such cluster will be equal to
$\mu_c \sqrt{\eta_{cj}}$ (in the case with string fusion),
where $\mu_c=\mu_0 \frac{\Delta S}{\sigma_0}=$
$\mu_0 \gamma^2=$
$\frac{d\mu}{dy}\Delta y \gamma^2$.
So to study the dependence on the cluster size $\Delta S$
in the given model we have to\\
1) increase the luminosity of elementary
emitters $\mu_0 \sqrt{\eta_{i}} \to \mu_c \sqrt{\eta_{cj}}=$
$\mu_0 \gamma^2 \sqrt{m_j/\gamma^2}=$ $\mu_0 \gamma \sqrt{m_j}$ \\
2) simultaneously reduce the number of clusters $M \to M_c=S(b)/\Delta S =$
$S(b)/(\sigma_0\gamma^2)=$ $M/\gamma^2$.

Note that there is no sense to introduce the clusters $\Delta S>\sigma_0$
(i.e. {\it correlated} fluctuations of the $\eta_i$ within area $\Delta S $)
in the case without string fusion.
Similarly it's no physical reasons to consider clusters
at small values of $\eta$  ($\eta<1$) even in the case with string fusion.

On the contrary at large $\eta$  ($\eta>>1$)
there are physical reasons in the case $\it with$ string fusion
to consider two possibilities:\\
A) $\Delta S=\sigma_0, \ \ \gamma^2=\Delta S/\sigma_0=1, \ \  M_c=M=S(b)/\sigma_0$\\
B) $\Delta S=S(b), \ \ \gamma^2=\Delta S/\sigma_0=M, \ \  M_c=1$ (one cluster).\\
The first one corresponds to the model, where colour string fields
are summing up only locally (formulated in section 2)
and the second one corresponds
to the model, where colour string fields are summing up
all over the interaction area $S(b)$ into one colour field.

In the case A) at the first stage
we have $M=S(b)/\sigma_0$ with $\eta_i$, $i=1,...,M$
fluctuated around $\eta$.
Then we have to generate particles from each area $\sigma_0$
with average multiplicities equal to $\mu_0 \sqrt{\eta_i}$
(see section 2).

{\bf Global string fusion.}
In the case B) we must MODIFY our formulae, as at first stage
we also have $M=S(b)/\sigma_0$ (like in the case A)) with $\eta_i$, $i=1,...,M$
fluctuated around $\eta$. Then (unlike the case A)) we have to find
average $\eta_c=\frac{1}{M}\seta=\frac{N}{M}$ for given event, and then to generate
particles from one cluster
with average multiplicity equal to $\mu_c \sqrt{\eta_c}=$
$\mu_0 M \sqrt{\eta_c}$ $=\mu_0 M \sqrt{N/M}$ $=\mu_0\sqrt{MN}$.
(Note that for configuration when all $\eta_i=\eta$ the average multiplicity
in the case A) and B) is the same $\mu_0 (S(b)/\sigma_0) \sqrt{\eta}$.
In general case they will be slightly different, as
$\sqrt{\frac{1}{M}\seta}$$\approx\frac{1}{M}\reta$.)

So in case B) we must replace (\ref{avO}) by
\beq
\avr{O} {\F}
=\frac{\sum_{\{\eta^{}_1,...,\eta^{}_M\}} \avr{O} {{\{\eta^{}_1,...,\eta^{}_M\}},\F} W(\Aeta)
p^{}_{\mu_c\sqrt{\frac{1}{M}\seta}} (\F)}
{\sum_{\{\eta^{}_1,...,\eta^{}_M\}} W(\Aeta)
p^{}_{\mu_c\sqrt{\frac{1}{M}\seta}} (\F)}
\label{avOcl}
\eeq
In the case B) we must ALSO MODIFY our expressions for
$\avr{O} {{\{\eta^{}_1,...,\eta^{}_M\}},\F}$ - the rates of the backward production from
configuration ${\{\eta^{}_1,...,\eta^{}_M\}}$. We have to use instead of (\ref{Onn})
for $\avr{\B}{\F}$ correlations:
\beq
\avr{\B} {{\{\eta^{}_1,...,\eta^{}_M\}},\F} =\mu_c \sqrt{\eta_c}= \mu_0 M \sqrt{\frac{1}{M}\seta}
\label{Onncl}
\eeq
with $M=S(b)/\sigma_0$ and instead of (\ref{Optn})
for $\avr{\ppB}{\F}$ correlations:
\beq
\avr{\ppB} {{\{\eta^{}_1,...,\eta^{}_M\}},\F} = p^2 \sqrt{\eta_c}=p^2 \sqrt{\frac{1}{M}\seta}
\label{Optncl}
\eeq
Again we see that the difference with the case A) consists in replace
$\frac{1}{M}\reta\to\sqrt{\frac{1}{M}\seta}$. As a consequence calculations
in the case B) are much more simple as we can reduce all sums $\sum_{\{\eta^{}_1,...,\eta^{}_M\}}$
to one sum $\sum_N$ using identity (\ref{1})
as in the no fusion case.

So in the fusion case with one cluster (case B))
we can write simple formulas as in the no fusion case.
Namely we have
\beq
\avr{\B}{\F}
=\frac{\mu_0 \sqrt{M}
\sum_N \sqrt{N} W(N) p^{}_{\mu_0\sqrt{M} \sqrt{N}}(\F)}
        {\sum_N W(N) p^{}_{\mu_0\sqrt{M} \sqrt{N}}(\F)}
\label{BFcl}
\eeq
and
\beq
\avr{\ppB}{\F}
=\frac{p^2}{\sqrt{M}}
\frac{\sum_N \sqrt{N} W(N) p^{}_{\mu_0\sqrt{M} \sqrt{N}}(\F)}
     {\sum_N W(N) p^{}_{\mu_0\sqrt{M} \sqrt{N}}(\F)}
\label{pBFcl}
\eeq
with $M=S(b)/\sigma_0$ and $W(N)$ is given by the formula (\ref{WN}).
We see that in this case B) (one cluster at large $\eta$)
n-n and pt-n correlations are connected
\beq
\avr{\ppB}{\F}=\frac{p^2}{\mu_0 M} \avr{\B}{\F}
\label{corcon}
\eeq

Note also that in the case A)
for Poissonian distributions $p^{}_{\eta_i}(n_i)$ (\ref{Poisson})
we have for $P^{}_{\{\eta^{}_1,...,\eta^{}_M\}} (\F)$ the formula (\ref{Poisso})
which is very similar to (\ref{avOcl}) but with
$\frac{1}{M}\reta$ instead of $\sqrt{\frac{1}{M}\seta}$.
More over we can fulfil in the case A)
the summation (\ref{PnF}) over ${\{n^{}_1,...,n^{}_M\}}$
and get the similar formula for $P^{}_{\{\eta^{}_1,...,\eta^{}_M\}} (\F)$
also in the cases of binomial and
negative binomial distributions (see the next section).
In these formulas the rate of $\F$ production depends only
on some average quantities for configuration ${\{\eta^{}_1,...,\eta^{}_M\}}$
(on $\frac{1}{M}\reta$ in the case A) and on
$\sqrt{\frac{1}{M}\seta}$ in the case B))
So we expect that we'll have very similar results for n-n and pt-n
correlations in the cases A) and B).

%  6  XXXXXXXXXXXXXXXXXXXXXXXXXXXXXXXXXXXXXXXXXXXXXXXXXXXXXX

\section{Non-Poissonian ( binomial, negative binomial ) distributions.}

{\bf Important note:}
Even if we have no Poisson distributions (\ref{nfpois})
for $p^{}_{\eta_i} (n_i)$ we still can get Gauss type formula (\ref{pgaus})
for $P^{}_{\{\eta^{}_1,...,\eta^{}_M\}} (\F)$ at $M\gg 1$ using (\ref{PnF}) and
the central limit theorem of the probability theory
\beq
P^{}_{\{\eta^{}_1,...,\eta^{}_M\}} (\F) \rightarrow  g^{}_{\av{\F}, \sigma_F}(\F)
\label{PnFgaus}
\eeq
with  $\av{\F}=\sum_i \bn_i$ and $\sigma_F^2=\sum_i \sigma_i^2$,
but now $\sigma_F^2$ can be not equal to $\av{\F}$.

Let us consider, as an example, the binomial distributions instead
of Poisson distributions.
\beq
p^{}_{\eta_i} (n_i) =
\beta^{}_{k_i, \lambda}(n_i)
\equiv C^{n_i}_{k_i}\lambda^{n_i}(1-\lambda)^{k_i-n_i}
\label{binom}
\eeq
Unlike Poisson distribution there are two parameters in binomial
distribution. We choose the same parameter $\lambda$ for all
$p^{}_{\eta_i} (n_i)$ distributions. In this case we can sum
these distributions like Poisson distributions due to formula
\beq
\beta^{}_{\ski, \lambda}(\F)
=\sum_{\{n^{}_1,...,n^{}_M\}} \delta_{\F,\sni} \prod_{i=1}^M  \beta^{}_{k_i, \lambda}(n_i)
\label{sbinom}
\eeq
For binomial distributions we have
\beq
\bn_i=k_i \lambda \hs{2}  \sigma_i^2=k_i \lambda (1-\lambda) =\bn_i (1-\lambda)
\label{parbinom}
\eeq
In the no fusion case we also have $\bn_i=\mu_0 \eta_i$ and hence
$k_i=$$\bn_i/\lambda=$$\frac{\mu_0}{\lambda} \eta_i$. So we have to choose $\lambda$
so that to ensure the $\frac{\mu_0}{\lambda}$ will be integer,
in other respects the parameter $\lambda$ is arbitrary.
(Clear that $n_i^{max}=k_i$ and the case of small $\lambda$, when
$k_i=n_i^{max} \gg \bn_i$, corresponds to the Poisson case.)
Then we find
\beq
P^{}_{\{\eta^{}_1,...,\eta^{}_M\}} (\F) =
\beta^{}_{\frac{\av{\F}}{\lambda}, \lambda}(\F)
\label{Binom}
\eeq
with  $\av{\F}=$$\sum_i \bn_i=$$\mu_0 \seta $ and
$\sigma_F^2=$$\sum_i \sigma_i^2=$$(1-\lambda) \mu_0 \seta $.
We see that for binomial distribution
$\sigma_F^2=(1-\lambda) \av{\F}$ with $0<\lambda<1$.
Note that case $\lambda\to 1$ corresponds to the situation,
when all $n_i=n_i^{max}=k_i$ and there are no dispersion
all $\sigma_i^2=0$.

Similarly for $w (\eta_i)$ starting instead of (\ref{wpois}) from
\beq
w (\eta_i) =  \beta^{}_{r_{i}, \lambda_\eta}(\eta_i)
\label{wbinom}
\eeq
with
\beq
\beti=r_{i} \lambda_\eta=\eta \hs{2}  \sigma_{\eta_i}^2
=r_i \lambda_\eta (1-\lambda_\eta) =\beti (1-\lambda_\eta)
\label{parwbinom}
\eeq
where now $\eta/\lambda_\eta$ must be integer.
Then we find instead of (\ref{WNpois})
\beq
W (N) = \beta^{}_{\frac{M\eta}{\lambda^{}_{\eta}}, \lambda_\eta}(N)
\label{WNbinom}
\eeq
with  $\bN=$$\sum_i \beti=$$M\eta $ and
$\sigma_N^2=$$\sum_i \sigma_{\eta_i}^2=$$(1-\lambda_\eta)M\eta $
and again
$\sigma_N^2=(1-\lambda_\eta) \bN$ with $0<\lambda_\eta<1$.
Clear that the case $\lambda_\eta\to 0$ corresponds to the Poissonian case and
the case $\lambda_\eta\to 1$ corresponds to the situation,
when $N=\bN$ and there are no dispersion $\sigma_N^2=0$ (fixed number of strings).

Then instead of (\ref{BF4}) we find
\beq
\avr{\B}{\F}
=\frac
{\mu_0\sum_N N
\beta^{}_{\frac{\bN}{\lambda^{}_{\eta}}, \lambda_\eta}(N)
\beta^{}_{\frac{\mu_0}{\lambda} N, \lambda}(\F)}
{\mu_0\sum_N
\beta^{}_{\frac{\bN}{\lambda^{}_{\eta}}, \lambda_\eta}(N)
\beta^{}_{\frac{\mu_0}{\lambda} N, \lambda}(\F)}
\label{BFbin1}
\eeq
In Gauss approximation we have the same formulas
(\ref{pgaus}) and (\ref{wgaus}) but now
with  $\sigma_F^2=\avr{\F}{N} (1-\lambda)=\mu_0 N (1-\lambda)$
and  $\sigma_N^2=\bN (1-\lambda_\eta)$.
We have also the same formula (\ref{BF5}) but now with
\beq
\varphi(N,\F)=\frac{(N-\bN)^2}{2\bN(1-\lambda_\eta)}
+\frac{(\F-\mu_0 N)^2}{2\mu_0 N(1-\lambda)}
\label{phibin}
\eeq
Clear that the case $\lambda\to 1$
corresponds to $\delta (\F-\mu_0 N)$ and $N=\F/\mu_0$.
Each string emits exactly $\mu_0$ particles and
the number of strings $N$ in the given event can be uniquely
reconstructed on the value of $\F$ which leads to 100\% correlations
(see discussion in the end of the section 4).
The case $\lambda_\eta\to 1$
corresponds to $\delta (N-\bN)$ and $N=\bN$,
which corresponds to the fixed number of strings.
There are no dependence of $N$ on the value of $\F$
in this case which leads to no correlations
(note that we consider the no fusion case).

We'll have all the same formulae (\ref{BF6},\ref{BF7},\ref{b1}) for $\avr{\B}{\F}$ and $b$,
but with modified equation for $z(f)$:
\beq
z^3-z^2=\frac{\mu_0\kappa}{2}(f^2-z^2)
\label{zbin}
\eeq
where
\beq
\kappa=\frac{1-\lambda_\eta}{1-\lambda}
\label{kappa}
\eeq
We see again that at $\lambda\to 1$ we have $\kappa\to\infty$ and $z=f$;
at $\lambda_\eta \to 1$ we have $\kappa\to 0$ and $z=1$ (doesn't depend on $f$).
Instead of (\ref{dz}) and (\ref{b2}) we have
\beq
\frac{dz}{df}=\frac{\mu_0\kappa f}{3z^2+z(\mu_0\kappa-2)}
\label{dzbin}
\eeq
and then
\beq
b=\frac{\mu_0\kappa}{\mu_0\kappa+1}
\label{b2bin}
\eeq
We see that really
at $\lambda\to 1$ we have $\kappa\to\infty$ and the correlation coefficient $b=1$;
at $\lambda_\eta \to 1$ we have $\kappa\to 0$ and the correlation coefficient $b=0$.

Clear that we can do all the same for negative binomial distributions.
We can also use any combinations of these distributions - one type
for $p^{}_{\eta_i} (n_i)$ and another type for $w (\eta_i)$.

%  7  XXXXXXXXXXXXXXXXXXXXXXXXXXXXXXXXXXXXXXXXXXXXXXXXXXXXXX

\section{Conclusion. Acknowledgments.}

We see that suggested simple cellular model
for the description of
the $p_t$ and multiplicity correlations
in high-energy nuclear collisions
and the Gauss approximation which
enables explicit analytical calculations
of the correlation functions in some asymptotic cases
give the adequate results in the no fusion case.

In the next paper we plan to present the results of
the correlation functions calculations
with taking into account the string fusion phenomenon,
based both on numerical summations on the configurations
$\Ceta$ using formulae of the section 2
and on analytical calculations using
the Gauss approximation discussed in section 4.

In conclusion the authors would like to thank M.A.~Braun
and G.A.~Feofilov for numerous valuable discussions.
The work has been partially supported by
the Russian Foundation for Fundamental Research
under Grant No. 01-02-17137-a.

%  Bibl  XXXXXXXXXXXXXXXXXXXXXXXXXXXXXXXXXXXXXXXXXXXXXXXXXXXXXXXX

%  Figs  XXXXXXXXXXXXXXXXXXXXXXXXXXXXXXXXXXXXXXXXXXXXXXXXXXXXXXXX

%%% BEGIN Fig.1%%%%%%%%%%%%%%%%%%%%%%%%%%%%%%%%%%%%%%%%%%%%%%%%%%
\begin{figure}[b]
% t-top, b-bottom, p-page, h-here
\centerline{\epsfig{file=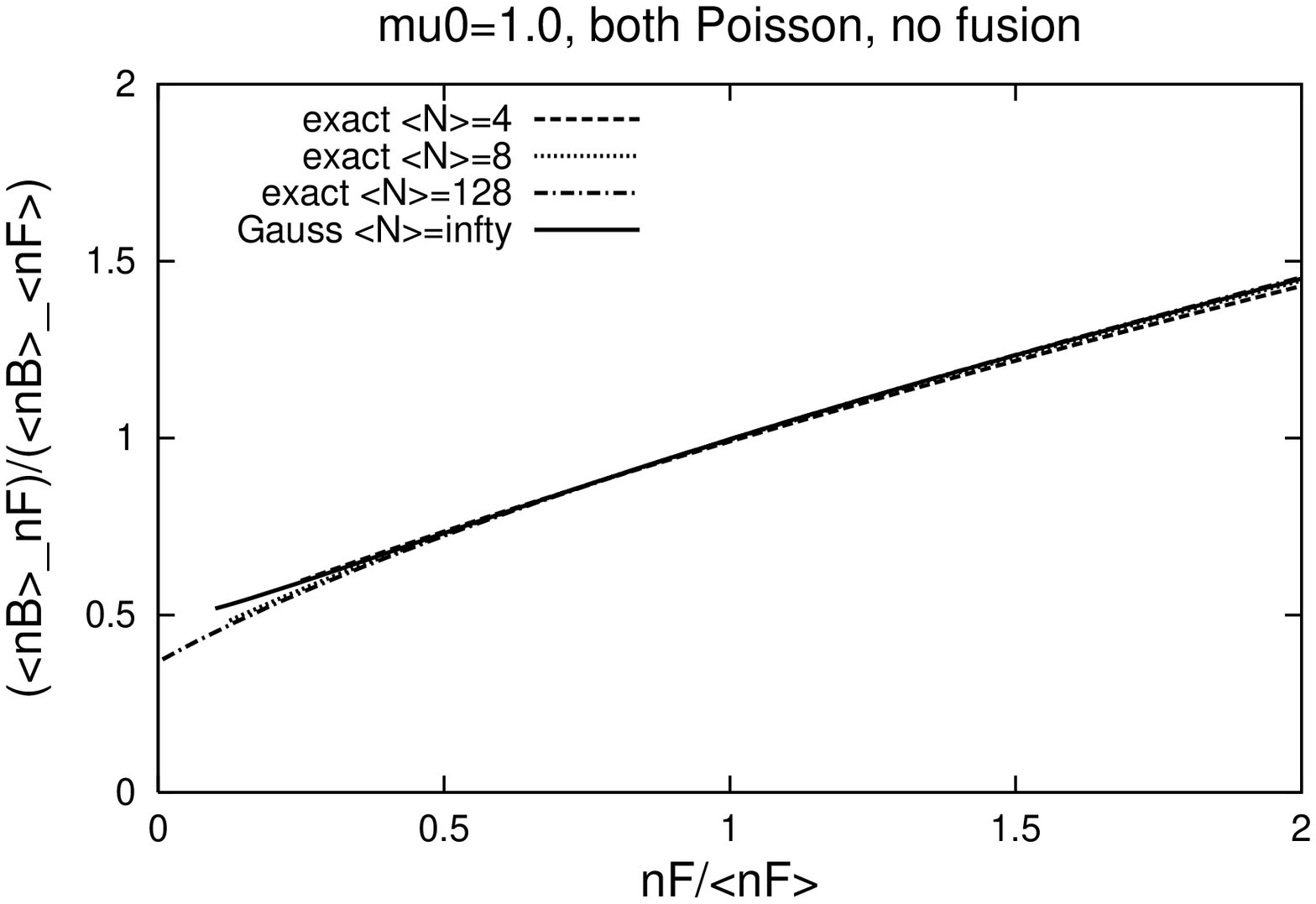,width=10cm,angle=-90}}
\caption[dummy]
{$\avr{\B}{\F}$ correlation functions in the no fusion case,
$\bN$-independence.
The check of validity of the Gauss approximation.
The $\av{N}=\bN$ is the mean number of strings,
the $\mu_0=\mu_{0F}$ is the mean number of particles
emitted by one string in the forward rapidity window, $\mu_0=1$.
(The $p(n_i)$ and $w(\eta_i)$ are both Poissonian.)
}
\label{vestn1_1}
\end{figure}
%%% END Fig.1%%%%%%%%%%%%%%%%%%%%%%%%%%%%%%%%%%%%%%%%%%%%%%%%%%%%
%%% BEGIN Fig.2%%%%%%%%%%%%%%%%%%%%%%%%%%%%%%%%%%%%%%%%%%%%%%%%%%
\begin{figure}[t]
% t-top, b-bottom, p-page, h-here
\centerline{\epsfig{file=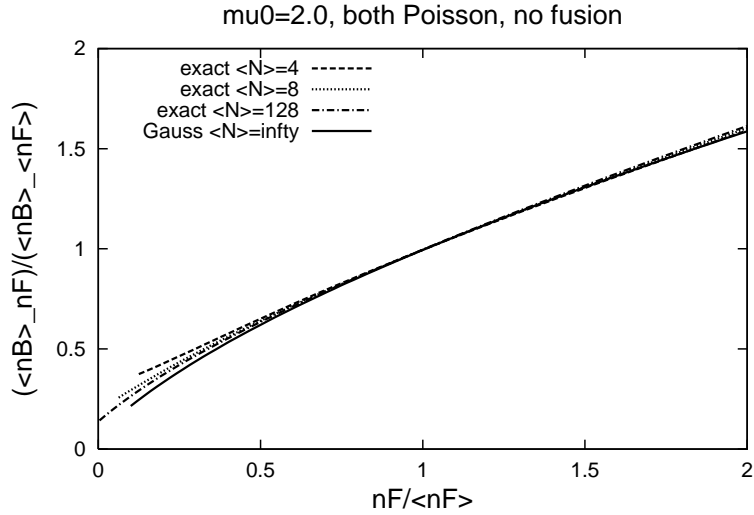,width=10cm,angle=-90}}
\caption[dummy]
{The same as in Fig.1, but for $\mu_0=2$.}
\label{vestn1_2}
\end{figure}
%%% END Fig.2%%%%%%%%%%%%%%%%%%%%%%%%%%%%%%%%%%%%%%%%%%%%%%%%%%%%
%%% BEGIN Fig.3%%%%%%%%%%%%%%%%%%%%%%%%%%%%%%%%%%%%%%%%%%%%%%%%%%
\begin{figure}[t]
% t-top, b-bottom, p-page, h-here
\centerline{\epsfig{file=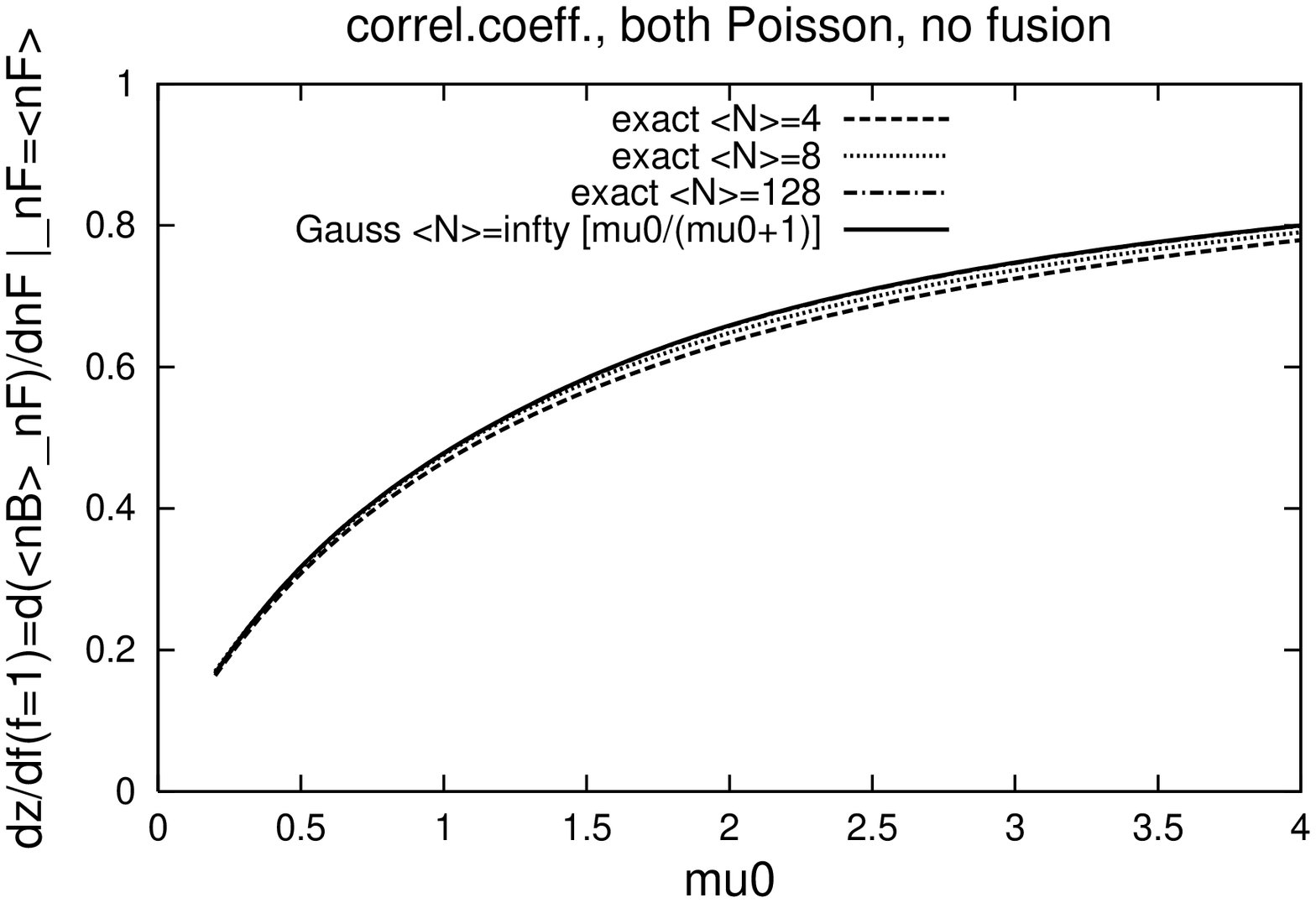,width=10cm,angle=-90}}
\caption[dummy]
{$\avr{\B}{\F}$ correlation coefficient $\olb$ (\ref{b4})
in the no fusion case as function of $\mu_0$,
$\bN$-independence.
The check of validity of the Gauss approximation.
The $\av{N}=\bN$ is the mean number of strings,
the $\mu_0=\mu_{0F}$ is the mean number of particles
emitted by one string in the forward rapidity window.
(The $p(n_i)$ and $w(\eta_i)$ are both Poissonian.)
}
\label{vestn1_3}
\end{figure}
%%% END Fig.3%%%%%%%%%%%%%%%%%%%%%%%%%%%%%%%%%%%%%%%%%%%%%%%%%%%%
%%% BEGIN Fig.4%%%%%%%%%%%%%%%%%%%%%%%%%%%%%%%%%%%%%%%%%%%%%%%%%%
\begin{figure}[t]
% t-top, b-bottom, p-page, h-here
\centerline{\epsfig{file=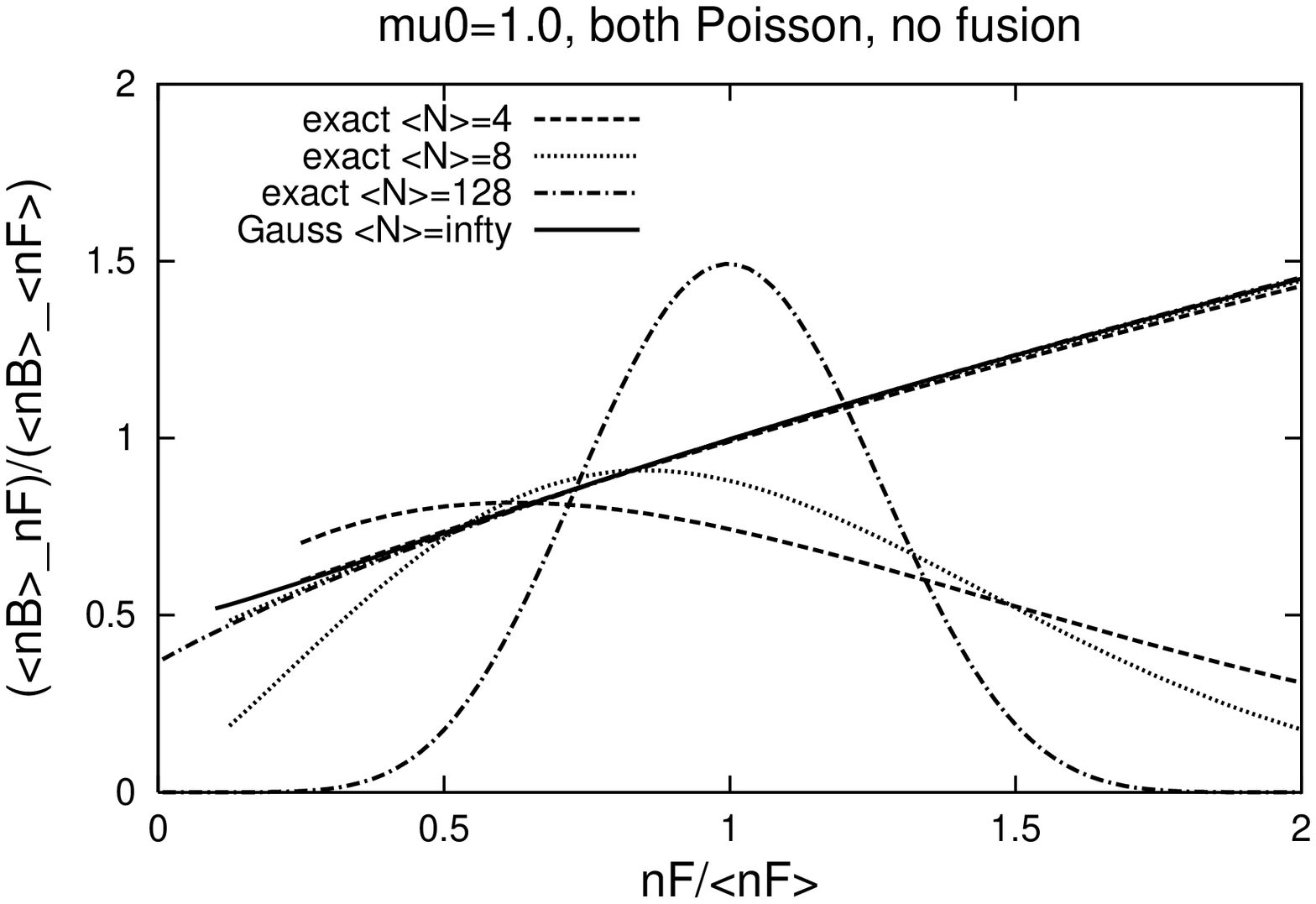,width=10cm,angle=-90}}
\caption[dummy]
{The same as in Fig.1, but plots for $P(\F)$ added
(not normalized, arbitrary units).}
\label{vestn1_4}
\end{figure}
%%% END Fig.4%%%%%%%%%%%%%%%%%%%%%%%%%%%%%%%%%%%%%%%%%%%%%%%%%%%%
%%% BEGIN Fig.5%%%%%%%%%%%%%%%%%%%%%%%%%%%%%%%%%%%%%%%%%%%%%%%%%%
\begin{figure}[t]
% t-top, b-bottom, p-page, h-here
\centerline{\epsfig{file=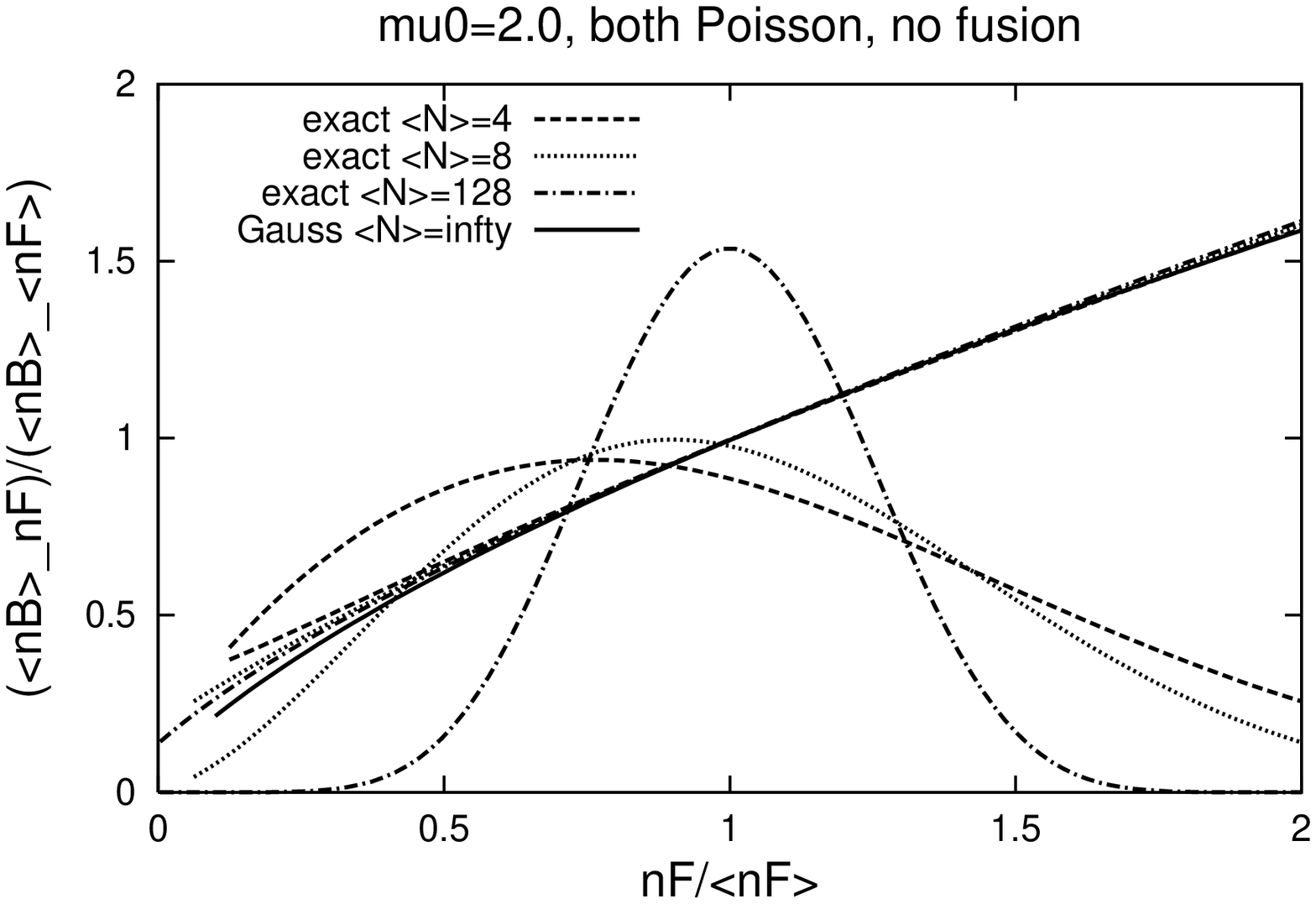,width=10cm,angle=-90}}
\caption[dummy]
{The same as in Fig.2, but plots for $P(\F)$ added
(not normalized, arbitrary units).}
\label{vestn1_5}
\end{figure}
%%% END Fig.5%%%%%%%%%%%%%%%%%%%%%%%%%%%%%%%%%%%%%%%%%%%%%%%%%%%%

\end{document}